Automated Reconstruction of Spherical Kikuchi Maps

Chaoyi Zhu[1], Kevin Kaufmann[2], Kenneth Vecchio[1,2,*]

[1]Materials Science and Engineering Program, University of California San Diego, La Jolla, CA 92093, USA

[2]Department of NanoEngineering, University of California San Diego, La Jolla, CA 92093, USA

Abstract

An automated approach to reconstruct spherical Kikuchi maps from experimentally collected electron backscatter diffraction patterns and overlay each pattern onto its corresponding position on a simulated Kikuchi sphere is presented in this study. This work demonstrates the feasibility of warping any Kikuchi pattern onto its corresponding location of a simulated Kikuchi sphere and reconstructing a spherical Kikuchi map of a known phase based on any set of experimental patterns. This method consists of the following steps after pattern collection: 1) pattern selection based on multiple threshold values; 2) extraction of multiple scan parameters and phase information; 3) generation of a kinematically simulated Kikuchi sphere as the 'skeleton' of the spherical Kikuchi map; and 4) overlaying the inverse gnomonic projection of multiple selected patterns after appropriate pattern center calibration and refinement. In the case study of pure aluminum, up to 90% of the Kikuchi sphere could be reconstructed with just seven experimentally collected patterns. The proposed method is the first automated approach to reconstructing spherical Kikuchi maps from experimental Kikuchi patterns. It potentially enables more accurate orientation calculation, new pattern center refinement methods, improved dictionary-based pattern matching, and phase identification in the future.

Keywords: electron diffraction, SEM, EBSD, Kikuchi band, spherical Kikuchi map, automated reconstruction, kinematic Kikuchi sphere simulation, inverse gnomonic projection

1. Introduction

Diffraction-based characterization techniques have been widely adopted to probe many aspects of structure and properties of materials. For instance, the powder X-ray diffraction technique is used to distinguish different phases based on reflected X-ray peaks' position (translation lattice variation) and intensity (crystal structure) (Bragg & Bragg, 1913), and neutron diffraction provides an accurate measurement of the lattice constant, which could be used to calculate residual stresses (Allen et al., 1985). Moreover, the development of electron diffraction-based techniques such as electron backscatter diffraction (Schwartz et al., 2000), transmission Kikuchi diffraction (Trimby, 2012), 3D EBSD (Zaafarani et al., 2006), electron channeling contrast imaging (Zaefferer & Elhami, 2014), electron imaging (Wright et al., 2015), transmission electron microscopy (Williams & Carter, 2009), etc. provide rich information including chemical composition, microstructure, grain boundaries, dislocation substructures, texture, residual stress/strain, and sub-angstrom atom arrangements.

Combined with energy dispersive X-ray spectroscopy (EDS), electron backscatter diffraction (EBSD) has now become a standard tool for laboratory work. Compared to other techniques utilizing a transmission electron microscope, EBSD offers much faster acquisition of a much larger sampling area and readily extracts information regarding microstructure (Humphreys, 2001), topography (Wright et al., 2015), crystallographic orientation (Schwartz et al., 2000), dislocation density (Demir et al., 2009; Zhu et al., 2018, 2017, 2016), texture (Engler & Randle, 2014), etc. Since its early discovery by Nishikawa and Kikuchi in 1928 (Nishikawa & Kikuchi, 1928), this high-angle Kikuchi diffraction technique has been further developed to obtain crystallographic information (Alam et al., 1954; Venables & Harland, 1973; Venables & Bin-Jaya, 1977). The emergence of commercially available EBSD systems is, to some extent, accredited to the development of fully automated image analysis methods to index diffraction patterns by groups from and University of Bristol (Dingley et al., 1987; Dingley, 1984), Risø national lab (Krieger Lassen, 1992) and Yale University (Adams et al., 1993; Wright & Adams, 1992). At present, the development of current EBSD algorithms or technologies is centered around a few key topics: 1) improving the accuracy and applicability of orientation determination; 2) expanding its capabilities, and 3) increasing the acquisition speed and obtaining a higher quality Kikuchi pattern.

The spatial resolution of EBSD is around 40 nm (Chen et al., 2011), which can be greatly improved to reach below 10 nm utilizing transmission based techniques (Trimby, 2012). The angular resolution of the Hough-based EBSD is typically around 0.5° to 1° (Ram et al., 2015; Humphreys, 2001, 1999; Wright, Nowell, & Basinger, 2011), although different methodologies have been developed to significantly improve angular resolution such as HR-EBSD (Wilkinson et al., 2006; Villert et al., 2009), pattern comparison method (Brough et al., 2006), iterative indexing (Thomsen et al., 2013), and the 3D-Hough transform (Maurice & Fortunier, 2008). In addition, another challenge of using Hough-based indexing is the fact that it relies on the image quality of Kikuchi patterns. Poor pattern quality images are difficult to index. A few groups have developed template matching or dictionary indexing approaches that aim to provide better indexing methods (Chen et al., 2015; Wilkinson et al., 2018; Foden et al., 2018; Nolze et al., 2018) to tackle many indexing challenges (Singh et al., 2018; Marquardt et al., 2017; Tanaka & Wilkinson, 2018). With the increase of computing power and implementation of a CMOS sensor, the new Symmetry EBSD system developed by Oxford Instruments is capable of capturing and indexing around 3000 patterns per second (Goulden et al., 2018). The boost in acquisition speed enables mapping of large samples (10s of square mm) and collecting more accurate statistics about a sample in a shorter time, which significantly enhances efficiency. Development of indirect detection of EBSP using an exposed back-illuminated CMOS sensor is also currently an area of research to obtain high-quality Kikuchi patterns (Wilkinson et al., 2013). Moreover, design of direct detector of EBSP has attracted much attention from researchers from different disciplines in University of Strathclyde (Vespucci et al., 2015) and CERN (Llopart et al., 2007, 2002), which has been demonstrated by vendors such as Thermo Fisher Scientific's tilt free EBD (Vystavěl et al., 2018) and EDAX's 'Clarity$^{TM}$' detector.

While current EBSD pattern analysis remains two dimensional, an insightful paper by Day (Day, 2008) suggests a paradigm shift into spherical image analysis using a spherical Kikuchi map. The development of practical applications of spherical Kikuchi maps is still a relatively unexplored area of research despite its early introduction. The method of using 3D-Hough transform (Maurice & Fortunier, 2008) in essence parameterize the 3D K-lines into 2D and increases the sharpness of local extrema in the Hough space by taking into account the hyperbolic shape of Kikuchi lines, an early demonstration of the value of thinking about indexing in 3D. Appropriate treatment of the hyperbolic character of Kikuchi bands, as a direct consequence of gnomonic projection of Kossel cones onto the phosphor screen, is particularly important for transmission Kikuchi diffraction to obtain reliable indexing in which the pattern center lies outside the collected pattern (Trimby, 2012). Moreover, Basinger *et al*. have developed a method to refine pattern center position by warping the Kikuchi pattern onto a sphere around the corrected pattern to preserve parallelism of the band edges, which can reduce phantom strain in HR-EBSD (Basinger et al., 2011). A recent paper submitted by Hielscher *et al*. (Hielscher et al., 2018) significantly contributes to this topic by developing spherical Radon transform and spherical cross-correlation for indexing patterns and demonstrates accuracy of less than 0.1°.

Unlike the emphasis of original work by Day (Day, 2008), which focuses on how spherical Kikuchi map may be used in the future, the focus of the present work is to elucidate the important steps of a novel automated method for accurate transformation of multiple flat Kikuchi patterns onto a simulated Kikuchi sphere using inverse gnomonic transformation, based on the orientation matrix and several other experimental parameters.  Although the image warping technique mentioned in the Day's work is a similar approach, we believe that it is critical to systematically establish an approach that uses all the 11 parameters (Day, 2008) for automated reconstruction of spherical Kikuchi map, which was not performed by Day.  In this study, the proposed method uses experimental EBSD patterns, from what might be considered a library of patterns of an individual phase, to automatically reconstruct a spherical Kikuchi map through inverse gnomonic projections.  The term 'library' used here may constitute a collection of patterns of one phase, which represent Kikuchi patterns obtained from different orientations, or maybe a collection of reconstructed experimental spherical Kikuchi maps of different phases.  This practical method has the potential to benefit the following research directions.  Since the interplanar angles and band shapes are better preserved on the sphere, one of the research areas that could result is the development of new indexing algorithms that determine the crystallographic orientation with improved accuracy to achieve higher sensitivity in quantifying deformation. For example, the reconstructed Kikuchi sphere from experimentally collected patterns could serve as a library for all possible patterns instead of the dynamically simulated patterns currently utilized in dictionary-based EBSD approaches (Chen et al., 2015). The proposed method is similar to the spherical cross-correlation approach (Hielscher et al., 2018) except that the 'master' spherical Kikuchi pattern is based on experimentally collected patterns rather than dynamically simulated spherical Kikuchi pattern. Since the reconstructed spherical Kikuchi map covers the entire range of possible orientations, which can be used to carry out cross-correlation type image matching on the sphere, to obtain the orientation without using the Hough-transform.

Since the reconstructed Spherical pattern could be overlaid onto a simulated Sphere in standard cartesian crystal frame, other potential applications including calibration of the EBSD setup such as pattern center position and sample orientation with respect to detector screen are also possible through iterative optimization approach.

2 Methodology

The automated reconstruction process for spherical Kikuchi maps involves the following stages: 1) selection of high fidelity EBSPs with random crystallographic orientations from a collected set, or library, of EBSPs of the same phase; 2) extraction of input parameters regarding the unit cell, pattern center, Euler angles, phosphor screen width/height, detector/sample tilt, and accelerating voltage; 3) generate the kinematical Kikuchi sphere; and 4) overlay the inverse gnomonic projection of EBSPs onto a simulated Kikuchi sphere. We chose herein to utilize and leave the simulated sphere in view to visually aid the reader in confirming our technique correctly reconstructs the Kikuchi sphere from simulated and experimentally collected EBSPs. This method is implemented using Matlab software.

2.1 Pattern Selection and Extraction of Input Parameters

Electron diffraction is sensitive to defects in materials, which strongly affects the coherency of scattering. Since EBSD is a surface technique, careful sample preparation of an undeformed/strain-free material is required prior to the scan. For example, final polish using colloidal silica, vibratory polishing, electropolishing or ion milling will significantly improve surface quality. For plastically deformed materials, the electron backscatter diffraction patterns are usually of relatively poor qualities, i.e., low band contrast (BC), meaning that the band edges are blurred (Wright, Nowell, & Field, 2011). The loss of sharpness in Kikuchi bands is also reflected in the drop of band slop (BS) values, which will adversely affect the accuracy of Hough transform to index patterns, i.e., higher mean angular deviation (MAD). Residual elastic stress, which changes the interplanar angle of bands, will also increase the MAD value. Moreover, a diffraction pattern generated from a grain boundary contains overlapping patterns from two differently oriented grains. Accurate reconstruction of a spherical Kikuchi map from experimentally collected EBSPs relies heavily on high-quality EBSPs that are accurately indexed. Therefore, pattern filtering based on BC, BS, MAD and disorientation values is necessary to exclude patterns with poor pattern quality and reduce the number of patterns collected with similar orientations. Selection of threshold values used for BC, BS, and MAD are usually user-defined and depend on the data set. Based on the BC, BS and MAD values of all the patterns in a scan, $30^{th}$ percentile for MAD and 60th percentile for BC/ BS have been chosen to as the first pass filtering to select high-quality patterns regardless of their orientations. Then, a disorientation table was generated using the Euler angles of the remaining patterns after filtering for BC, BS, and MAD. By applying the symmetry operators to the pairs of misorientation calculations (Heinz & Neumann, 1991), the minimum value of misorientations is determined as physically plausible misorientation angles within the fundamental zone i.e. the disorientation angles. A reasonable threshold value for the disorientation angle is the high-angle grain boundary (~15˚), which enables selection of patterns from significantly different orientations. After pattern filtering, a pre-selected group of differently orientated patterns can be obtained, which has good quality and relatively better absolute orientation accuracy. Since the MAD values does not effectively reflect the absolute orientation accuracy, an additional selection metric is introduced. The alignment of the experimental patterns with kinematically simulated pattern is examined through 2D normalized cross-correlation as a measure of goodness of match prior to reconstruction (Winkelmann et al., 2016). Selection criteria can be subsequently established based the highest normalized cross-correlation coefficients in order to produce well-aligned spherical Kikuchi map.

The reconstruction technique also requires patterns representing crystallographically-distinct orientations to fill as much space on the sphere as possible, which is feasible in a polycrystalline material with moderately random texture. Through crystal symmetry, equivalent Euler angle sets (different Euler angle values but same Kikuchi diffraction) can also be calculated (Nolze, 2015) to simplify the current approach and reconstruct the 'missing' areas. Other parameters such as detector tilt, pattern center, accelerating voltage and pattern width/height should also be accessible either through the EBSD software or stored as image metadata. For the new Aztec-HKL software from Oxford instruments, these parameters are stored as image metadata. Experimental EBSD patterns were collected utilizing an Oxford Symmetry detector in Resolution mode (1244 by 1024 pixels per image). These experimental patterns serve as a library of experimentally collected EBSD patterns from which we reconstruct the Kikuchi sphere for that material.

2.2 Coordinate Transformations in EBSD

A thorough description of any crystal plane/direction relative to the Kikuchi pattern on the detector requires knowledge of all the relevant coordinate systems. The five (right-handed) coordinate systems illustrated in Fig. 1 is listed in the following:

1. the phase-specific **crystal lattice frame** (**a,b,c**), which conveniently defines a crystal through translation of unit cells and simplifies crystallographic calculations.

2. the orthogonal **cartesian crystal frame** ($X_c$, $Y_c$, $Z_c$) redefines the coordinates of atom positions in Euclidean space for metric calculations.

3. the **sample frame** ($X_s$, $Y_s$, $Z_s$), which sits at the beam position and describes the tilted sample coordinate system. In the Oxford Instruments and Bruker Nano EBSD systems, this sample frame also represents the reference coordinate system for Euler angle definition. In the EDAX TSL EBSD system, it is rotated around the $Z_s$ axis by 90°, which results in a 90° offset of the Euler angle.

4. the **detector frame** ($X_d$, $Y_d$, $Z_d$), which coincides with the origin of the sample frame at the beam position and its $Z_d$ is normal to the detector screen.

5. the **2D gnomonic projection frame** (**X₉, Y₉**), which is centered at the pattern center on the detector screen.

The transformations used to relate crystallographic planes/directions and Kikuchi patterns will be briefly covered, similar to a more comprehensive work by Britton *et al.* (Britton et al., 2016). Starting with a crystal direction denoted as row vector [u,v,w] in crystal lattice frame, it is necessary to first describe its position in the orthogonal cartesian crystal frame. Choice of cartesian crystal frame in this study follows the convention that: 1) it aligns the **c** axis with **Z_c**, and 2) the **b** axis is in the **Y_c-Z_c** plane to maintain consistency. It is known that the six parameters about the unit cell are: a = |**a**|, b = |**b**|, c = |**c**|, α = ∠(**b,c**), β = ∠(**a,c**), γ = ∠(**a,b**). The complete description of the unit cell can then be used to calculate the structure matrix A (Britton et al., 2016).

The row vector in cartesian crystal frame [u_c,v_c,w_c]_c is hence obtained by multiplying the transpose of the structure matrix $A^T$.

$$[u_c, v_c, w_c]_c = [u, v, w]A^T \qquad (1)$$

Following the Bunge convention in describing crystallographic orientation (ZXZ sequence), the cartesian crystal frame can be aligned with the sample frame through three 'passive' rotation matrices defined by the three Euler angles ($\phi_1$, $\Phi$, $\phi_2$). For instance, $R_{Z(\phi_1)}$ represents rotation around Z axis by an angle of $\phi_1$ and sample title around X axis is given by $R_{X(\theta_{tilt})}$. The orientation matrix O is the produce of three rotation matrices $R_{Z(\phi_2)}R_{X(\Phi)}R_{Z(\phi_1)}$.

The overall coordinate transformations that takes a crystal direction from the crystal lattice frame to the detector frame is given by the following equation.

$$[u_d, v_d, w_d]_d = [u, v, w]A^T OR_{X(\theta_{tilt})} \qquad (2)$$

Transformation of a crystal plane (hkl) is similar to a crystal direction except that it is initially transformed into the **reciprocal** cartesian crystal frame.

$$[h_d, k_d, l_d]_d = [h, k, l]A^{-1} OR_{X(\theta_{tilt})} \qquad (3)$$

2.3 Generation of Kinematical Kikuchi Sphere

A kinematically simulated Kikuchi sphere is produced in this study as the 'skeleton' of the reconstructed Kikuchi sphere, mainly for visualization purpose and pattern selection. It validates the accuracy of the reconstruction process since it provides a 'true' location of the experimental EBSP. Generation of a kinematical Kikuchi sphere includes computation of angles for all of the possible Bragg reflectors as well as intensities of the reflectors. A simple geometrical model can be used to describe the Bragg reflection from Friedel pairs of lattice planes with normals (hkl) and ($\bar{h}\bar{k}\bar{l}$). The *n*>1 case represents higher-order interferences, indicating that electrons are diffracting from the set of planes with spacing d/*n*, which are manifested as weak diffraction lines parallel to the *n*=1 Kikuchi diffraction lines in the kinematically simulated and experimentally collected EBSPs (Williams & Carter, 2009). The width of a Kikuchi band is given by approximately twice the Bragg angle $\theta_{hkl}$.

$$\sin \theta_{hkl} = \frac{n\lambda}{2d_{hkl}} \qquad (4)$$

Determination of the kinematical intensities of a Kikuchi band means to find the modulus squared of the corresponding structure factor. Based on the knowledge of the atomic scattering factor, the structure factor is computed to describe the collective scattering power from a unit cell. In a single-atom scattering model, the scattering power of the atom is the Fourier transform of its electrostatic potential based on the first Born approximation. X-ray (electromagnetic radiation) scattering is determined by its charge distribution and is treated as a kinematical (single) scattering process with tabulated atomic scattering parameters (Smith & Burge, 1962; Doyle & Turner, 1968). However, electron diffraction is a complex phenomenon since the scattering of high-energy electrons (charged particles that travel like waves) is a dynamic (multiple) scattering process (Zaefferer & Elhami, 2014). A kinematical scattering assumption is used in this study for simulating the Kikuchi sphere, which poorly simplifies the electron diffraction physics. A comprehensive discussion on the many limitations of the kinematical model can be found in (Winkelmann et al., 2016). Fortunately, studies have been done that deal with the dynamical scattering of electrons (Winkelmann, 2010; Callahan & De Graef, 2013) to simulate EBSPs.

Since the electrostatic potential function of atoms can be correlated with the charge density function through Maxwell's equation, the atomic scattering factor of electrons $f^{el}$ can be kinematically approximated using the X-ray scattering factor through the Moth-Bethe formula.

The structure factor of the unit cell is, therefore, a superposition of the single electron scattering from every atom of the unit cell:

$$F_{hkl} = \sum_{j=1}^{N} f_j^{el}(s)\, e^{2\pi i g_{hkl} \cdot r_j} \tag{5}$$

The kinematical intensity ($I_{hkl}$) is then given by $|F_{hkl}|^2$ in the absence of anomalous absorption effects. However, the actual intensity value from the dynamical scattering theory differ significantly from the kinematical intensity. A study by Winkelmann shows that the linear approximation (i.e. $I_{hkl} = |F_{hkl}|$) demonstrate closer looking Kikuchi band intensity through cross-correlation despite its apparent limitations (Winkelmann, 2010). In this study, a linear approximation of intensity is adopted.

On the simulated Kikuchi sphere, the coordinates of points on the Kikuchi sphere are represented in the cartesian crystal frame $[x_{ks}, y_{ks}, z_{ks}]_c$. Points on the Kikuchi sphere located within the Bragg's reflection angle of a reflector are assigned with intensity value $I_{hkl}$ of the corresponding reflector (hkl). A repetitive loop over all the reflectors will yield total intensity values of every point on the Kikuchi sphere due to the presence of all the reflectors.

$$I(x_{ks}, y_{ks}, z_{ks}) = \sum_{i=1}^{R} I_{hkl}^{i} \tag{6}$$

$$\begin{cases} if\ |[x_{ks}, y_{ks}, z_{ks}]_c \cdot [h_c^i, k_c^i, l_c^i]_c| \leq \sin(\theta_{hkl}^i),\ I_{hkl}^i = |F_{hkl}|, \\ \quad\quad\quad \text{otherwise}, \quad I_{hkl}^i = 0 \end{cases}$$

It is important here to note that the crystal plane coordinates $[h_c, k_c, l_c]_c$ in the cartesian crystal frame are obtained by post-multiplying the Miller indices [h,k,l] with the inverse of structure matrix A. Every point residing on the kinematically simulated Kikuchi sphere and its corresponding final intensity value $I(x_{ks}, y_{ks}, z_{ks})$ are then saved for later use. An example of the spherical Kikuchi lines and the kinematically simulated Kikuchi sphere are shown in Fig.2. If the purpose were to align simulated bands to an experimental pattern, the experimentally determined orientation matrix O and tilt matrix $R_{X(\theta_{tilt})}$ defined in Eq. 3 would be used to rotate the Kikuchi sphere into the correct orientation. In this study, the orientation of the Kikuchi sphere is invariant in the detector frame, meaning that the orientation and sample tilt matrices are identity matrices. As described in the next part, the EBSP patterns are then rotated in order to be mapped onto the correct location on the Kikuchi sphere.

2. 4 Inverse Gnomonic Projection

The projection of the backscattered electrons to the detector screen is geometrically defined by the gnomonic projection of the Kikuchi sphere from the beam position (center of the sphere at which all the red arrows begin), illustrated in Fig. 3. In other words, the intensity of points on the Kikuchi sphere is gnomonically translated onto a plane tangent to the sphere. In the current experimental setup, the detector orientation with respect to the sample is fixed, whereas the orientation of the actual Kikuchi sphere is changing depending on the orientation of the crystal relative to the primary incoming beam of electrons. Each EBSP collected on the detector is only a section of the Kikuchi sphere's surface that is in view of the detector screen. In practice, the distortion generated from the optics of the detector needs to be corrected prior to reconstruction (Day, 2008) unless it has already been corrected in the software. Furthermore, accurate knowledge of the position of each pattern center is also critical to accurately warp the Kikuchi pattern onto the Kikuchi sphere (Basinger et al., 2011). Pattern center calibration (Krieger Lassen, 1999; Carpenter et al., 2007; Maurice et al., 2011) using a calibration sample such as single crystal silicon is recommended, which should have already been completed during installation of the EBSD detector.

From the perspective of the reference systems introduced in 2.2, the origin of the detector frame coincides with the center of the Kikuchi sphere and the **Z_d** axis extends from the beam position and perpendicularly intersects with the detector at a point, the pattern center, which is colored in green as shown in Fig. 3. A new 2D coordinate gnomonic projection frame (**X_g, Y_g**) is assigned at the pattern center in order to accurately associate experimentally collected band positions with simulated bands. The mathematical description of the gnomonic transformation from points on Kikuchi sphere in detector frame (**X_d, Y_d, Z_d**) to points on EBSP in the 2D gnomonic projection frame (**X_g, Y_g**) is given as:

$$x_g = \frac{x_d}{z_d} \; ; \; y_g = \frac{y_d}{z_d} \tag{7}$$

Due to the assignment of the gnomonic projection frame, the pattern center position is ($x_g = 0$, $y_g = 0$). In the Oxford Instruments AZtec software, the pattern center (PC_X, PC_Y, DD) information available is a set of normalized unitless quantities, which can be defined in the following way:

- PC_X: horizontal position of pattern center, normalized to pattern width ($w_p$), measured from the left edge of Kikuchi pattern.

- PC_Y: horizontal position of pattern center, normalized to pattern width ($h_p$), measured from the bottom edge of Kikuchi pattern.

- DD: detector distance (closest distance from beam position to detector screen) normalized to pattern width ($w_p$).

All points on the Kikuchi sphere with $z_d = 0$ therefore lies infinitely away from the pattern center regardless of the value of $x_d$ or $y_d$. After appropriate transformation into the gnomonic coordinates, the tangent point (pattern center) is at a unit distance from the beam position, i.e., the Kikuchi sphere in the gnomonic projection frame is a unit sphere. The corner positions of the Kikuchi pattern image including the bottom left corner $[x_g^{min}, y_g^{min}]$ and top right corner $[x_g^{max}, y_g^{max}]$ can be determined according to eq.9 and eq.10 (Note that the gnomonic coordinate positions is a unitless quantity).

$$a_p = \frac{w_p}{h_p} \tag{8}$$

$$y_g^{min} = -\frac{PC_y}{DD}\frac{1}{a_p} \qquad y_g^{max} = +\frac{1 - PC_y}{DD}\frac{1}{a_p} \tag{9}$$

$$x_g^{min} = -\frac{PC_x}{DD} \qquad x_g^{max} = +\frac{1 - PC_x}{DD} \tag{10}$$

Given the pattern center position, each pixel on the EBSP (pixel location is I × j) can be assigned with gnomonic projection coordinates located within the corner positions. For a 2D gnomonic projection coordinates of $[x_g^i, y_g^j]$, the corresponding coordinates on the Kikuchi sphere of this point $[x_d^i, y_d^j, z_d^k]$ can be found using the following transformation rule (note that the value of $z_d^k$ is determined prior to $x_d^i$ and $y_d^j$).

$$[x_\mathbf{g}^i, y_\mathbf{g}^j] \rightarrow [x_\mathbf{d}^i = x_\mathbf{g}^i \cdot z_\mathbf{d}^k, y_\mathbf{d}^j = y_\mathbf{g}^j \cdot z_\mathbf{d}^k, z_\mathbf{d}^k = \sqrt{1 - (x_\mathbf{g}^i)^2 - (y_\mathbf{g}^j)^2}] \qquad (11)$$

In a repetitive inverse gnomonic projection process for all the patterns, the orientation of the kinematically simulated sphere is kept invariant in the cartesian crystal frame. Considering the tilt angle between the sample surface and the detector plane, as well as the orientation of the crystal, the coordinates for EBSP in the detector frame $[x_\mathbf{d}^i, y_\mathbf{d}^j, z_\mathbf{d}^k]$ can be rotated into the position from which the pattern is generated on the simulated Kikuchi sphere in the cartesian crystal frame by post-multiplying $(OR_{X(\theta_{tilt})})^{-1}$.

$$[x_\mathbf{ks}^i, y_\mathbf{ks}^j, z_\mathbf{ks}^k]_c = [x_\mathbf{d}^i, y_\mathbf{d}^j, z_\mathbf{d}^k] \, (OR_{X(\theta_{tilt})})^{-1} \qquad (12)$$

2.5 Alternative Rotation Methods

By convention, the orientation of crystal with respect to the sample reference frame is described by Euler angles in most of the commercial EBSD software despite the redundancy. Therefore, this study describes the rotation of the collected patterns using the orientation matrix calculated from three Euler angles and the sample tilt matrix, see Eq. 12.

It is also possible to use alternative methods based on the axis-angle pair to rotate the pattern. For instance, the most convenient method is to determine the Rodrigues' vector. Since axis-angle pair description is not directly available, extraction of axis-angle pair from combined orientation matrix and the sample tilt matrix would allow simple demonstration for using alternative rotation methods. Eq.13 shows the total rotation matrix for rotating the pattern on the Kikuchi sphere due to the orientation of the crystal and tilt of the sample.

$$R_{total} = OR_{X(\theta_{tilt})} \tag{13}$$

The axis-angle pair descriptor of the rotation is represented by the rotation axis **k** and the rotation angle $\theta$ around this rotation axis k using right hand rule. The rotation axis $\mathbf{k} = [k_x, k_y, k_z]$ of the rotation matrix $R_{total}$ is the eigenvector when the corresponding eigenvalue is equal to unity. The rotation angle (in radians) is determined through the following equation:

$$\theta = \text{acos}\left(\frac{tr(R_{total}) - 1}{2}\right) \tag{14}$$

The rotation of the position vector **v** in the detector frame i.e. $v_d = [x_d^i, y_d^j, z_d^k]$ into the cartesian crystal frame i.e. $v_c = [x_{ks}^i, y_{ks}^j, z_{ks}^k]_c$ can therefore be equivalently achieved through the axis-angle pair rotation.

$$v_c = v_d \cos\theta + (k \times v_d)\sin\theta + k(k \cdot v_d)(1 - \cos\theta) \tag{15}$$

Furthermore, the axis-angle pair notation can be further parametrized into more elegantly defined quaternions (W.R. Hamilton, 1899).

$$(q_0, q_1, q_2, q_3) = (\cos\frac{\theta}{2}, k_x \sin\frac{\theta}{2}, k_y \sin\frac{\theta}{2}, k_z \sin\frac{\theta}{2}) \tag{16}$$

The rotation of the pattern can be similarly written in terms of quaternions:

$$v_c = v_d + 2q_0(\omega \times v_d) + 2(\omega \times (\omega \times v_d)) \tag{17}$$

where $\omega = [q_1, q_2, q_3]$.

If axis-angle pair descriptor were directly given by commercial EBSD vender, the rotation of pattern could be carried out simply using the above two methods see Eq.15 and Eq.17. It has been verified in this study that the reconstructed Spherical Kikuchi pattern using the axis-angle pair or quaternions method is the same as the orientation matrix method. However, the added advantage is that the crystal orientation description using the axis-angle pair or quaternions is more robust compared to Euler angles.

3. Results and Discussion

3.1 Spherical Kikuchi Map Generated from Dynamically Simulated Patterns

A series of validation experiments are carried out using dynamically simulated patterns generated from Brukers' ESPRIT DynamicS (trial version) for austenite (fcc. Six Kikuchi patterns of random orientations are simulated for each material. The accelerating voltage is 20 kV and the pattern center positions are all (0.5,0.5,0.5). No sample tilt is involved in generating these patterns. Since every simulated pattern is representative of a high-quality unstrained pattern, no prior pattern filtering is needed. The top row of Fig. 4 shows unit cell structures of the four materials used in the validation experiment. The unit cell description is an important part of the input parameters for generating a Kikuchi sphere. A list of all possible crystal planes is generated before structure factor calculation. Based on calculated structure factors, a corresponding kinematical Kikuchi sphere is generated for each material at 20kV. The intensity values on the simulated Kikuchi sphere and simulated pattern are normalized with respect to the maximum intensity value. While keeping the orientation of the simulated Kikuchi sphere fixed, simulated Kikuchi maps are mapped onto the simulated Kikuchi sphere using the input Euler angles used to generate simulated patterns. It is therefore not surprising to observe that the Kikuchi sphere has all the band positions that almost perfectly align with the dynamically simulated pattern. As expected, the intensity values differ significantly between the two, since dynamically simulated patterns demonstrate more realistic intensity distribution (Winkelmann et al., 2016). An example of the reconstruction process is shown for austenite in Fig. 4. Before visualization, there is no interpolation involved in processing each of the pattern image intensities. The intensity data is stored in a matrix of the same dimension of the pattern image. However, the visualization of the entire reconstructed spherical involves storing the intensity values of multiple patterns into a regular grid. The off-grid intensity values are obtained through interpolation to achieve smoothly displayed spherical Kikuchi map. Currently, no image stitching technique has been employed. The purpose of doing reconstruction using dynamically simulated patterns is to verify the projection method as well as the off-grid cubic interpolation for storing the spherical Kikuchi map data applied on multiple images before using the experimentally collected patterns.

3.2 Spherical Kikuchi Maps Generated from Experimental Patterns

Reconstruction of experimental spherical Kikuchi maps is slightly more complicated. In practice, optics distortion from the detector system imposes barrel distortion to the edge of pattern images (Day, 2008). In Oxford Instrument's new Symmetry detector and AZtec software, distortion due to the optics has already been corrected. Hence, the experimental patterns used in this study are as-received patterns from the AZtec software.

Magnetic field distortion will result in curved Kikuchi bands, which should be corrected prior to Spherical Kikuchi map reconstruction. However, the magnetic field distortion introduced by the strong immersion lens used in SEM to focus the electron beam is only present in a few microscopes other than the Thermo-Fisher Apreo SEM used in this study. A special edition of the AZtec does implement a magnetic field correction method prior to band detection for those SEMs that require it (Chou et al., 2013). Those magnetic field distortion corrected pattern should be used with care for reconstruction since these patterns contain regions of no pattern information. In addition, the effect of sample charging on EBSPs should be avoided by mounting samples in the conductive mount, using silver paint or even running in low vacuum mode by introducing ionized water vapor.

First, pattern selection and extraction of experimental parameter are carried out, which has been described in section 2.1. The metadata stored in each pattern image is extracted and passed into the reconstruction code as an input parameter. The selected patterns are next reconstructed onto the simulated Kikuchi sphere one by one. The test materials used in this study include aluminum (fcc), tungsten (bcc), and tungsten carbide (hexagonal) synthesized in the lab with random texture as shown in Fig. 5. Fully reconstructed spherical Kikuchi maps for these selected materials are all shown in Fig.5. For aluminum and tungsten, it is sufficient to fully reconstruct the displayed hemisphere with patterns from different orientations. By employing the point reflection of the crystal symmetry, the other un-displayed hemisphere can also be reconstructed. In the case of tungsten carbide, the six-fold symmetry has been utilized to fully reconstruct the entire hemisphere by subtracting or adding multiples of 60° to the $\phi_2$ Euler angle as shown in S.3 in the Appendix (the fully reconstructed hemisphere displayed in the Appendix requires -60° and 60° rotation of the original patterns). The kinematically simulated sphere is utilized for the reader's benefit; to visually confirm that the reconstruction of experimentally collected EBSPs onto the Kikuchi sphere is valid. The reader can now easily see the experimentally collected

Kikuchi bands aligning with their simulated counterpart. The reconstruction process for aluminium, tungsten and tungsten carbide can be found in the Appendix (Figures S1-3).

3.3 Applications and limitations

In the original paper by Day (Day, 2008), several potential applications have been identified: 1) automatic reconstruction of spherical Kikuchi map (SKM); 2) new indexing and band detection algorithms; 3) strain measurement of polycrystalline materials; 4) phase identification of unknown phases. This study specifically deals with the first suggestion, which establishes a systematic approach to precisely overlay experimentally collected Kikuchi patterns onto a simulated Kikuchi sphere for potential use in the following suggested applications as well as other techniques. As mentioned previously, new indexing approaches such as spherical cross-correlation and other various dictionary-based indexing approaches (Foden et al., 2018; Chen et al., 2015; Nolze et al., 2018) rely on the use of a dynamically simulated spherical Kikuchi map (Hielscher et al., 2018). The template matching or dictionary indexing approach requires simulation of the enormously large and sparse library of patterns of different orientation. Therefore, matching of exact orientation is practically impossible since the generated library requires 'steps' to be set for Euler to produce a finite library. The advantage of using the spherical cross-correlation method compared to a dictionary approach is that it directly looks for a location on a dynamically simulated Kikuchi sphere to best match the experimental pattern and then computes the associated rotation matrix. However, the simulated Kikuchi sphere rarely represents the real intensities and Kikuchi band profiles observed in the Kikuchi pattern. Using spherical Kikuchi map reconstructed from experimentally collected patterns, it represents many realistic intensities of Kikuchi bands and allows spherical cross-correlation to be carried out. Hence, it will further reduce computation cost and possibly increase the accuracy of orientation measurement.

Although not specifically mentioned in Day's paper (Day, 2018), the accuracy of pattern center determination drives the accuracy of the reconstruction, which suggests this could be exploited as a new method for pattern center refinement with a relative PC error of 0.045% (Basinger et al., 2011). Since one of the limiting factors in Basinger *et al.*'s work is the initial orientation measurement error, PC calibration based on multiple patterns collected from different orientations on a single spherical Kikuchi map might be able to mitigate this problem. Similarly, the reconstruction could also be used to obtain more accurate detector to sample tilt angle $\theta_{tilt}$ based on many reconstructed patterns, which should reduce systematic misalignment between the simulated Kikuchi sphere and all reconstructed experimental patterns according to eq. 12. Moreover, although lens correction could be more conveniently addressed by imaging an evenly spaced grid, it is possible to utilize the parallel nature of Kikuchi bands to iteratively converge to a lens aberration correction function.

Reconstructing an 'undeformed' spherical Kikuchi map using experimental patterns collected from polycrystalline material is difficult since a theoretically unstrained polycrystalline sample is practically non-existent. Sample preparation and heat treatment all needed to be meticulously carried out to really obtain a 'strain free' surface of a polycrystalline material. For a deformed polycrystalline material, the distribution of residual strain is highly heterogeneous, which makes the reconstruction for a particular strain level nearly impossible. The practical aspect for conducting strain measurements on the Kikuchi sphere is to adopt the idea of obtaining absolute strain by comparing experimental and simulated patterns (Kacher et al., 2009; Vermeij et al., 2019; Villert et al., 2009), but on a sphere using spherical cross-correlation. However, this technique still faces many more challenges to achieve its desired sensitivity for strain mapping since it requires extremely accurate knowledge of many experimental parameters (Britton et al., 2010).

In the current study, the identity of each phase utilized is known before it is inserted into the microscope. The state-of-the-art EBSD technology does couple with EDS data to incorporate a phase identification function, to assist in identifying unknown phases. Similar to ab-initio analysis on diffraction patterns developed for CBED (Ayer, 1989; Page, 1992), phase identification using just EBSD patterns (Dingley & Wright, 2009; Han, Chen, et al., 2018; Li et al., 2014; Li & Han, 2015; Michael & Eades, 2000; Michael & Goehner, 1999; Kaufmann et al., 2019), chemical-sensitive holography (Lühr et al., 2016) or in combination with EDS data (Nowell & Wright, 2004; Small & Michael, 2001) have also been extensively studied to compete with the more traditional XRD method. Due to the inherent nature of diffuse scattering, the accuracy of any electron diffraction-based method to measure lattice parameters is unlikely to reach that of XRD without a sophisticated band localization algorithm (Ram et al., 2014), although correct classification of the Bravais lattice with a reasonably accurate lattice parameter is already possible (Han, Zhao, et al., 2018; Han, Chen, et al., 2018; Michael, J. R., & Goehner, 2000; Michael & Goehner, 1999). In order to determine the symmetry elements from EBSPs, (e.g. rotation axes, diads, triads, etc.), correction for distortion caused by gnomonic projection and lens aberration are necessarily carried out. Another potential application of the spherical Kikuchi map would be to carry out dictionary-based phase identification. For example, an experimental diffraction pattern could be used to search for potential candidates from a library of simulated master Kikuchi spheres of different phases and select, based on a comparison of multiple patterns, the phase or list of potential phases with an optimal match. Furthermore, the reconstruction approach is not just limited to EBSD, but should work for reconstruction of convergent beam electron diffraction (CBED) patterns obtained in a TEM as well given that we know the zone axis and the geometry of the diffraction experiments, which could even assist in distinguishing different space groups in the SKM dictionary-based *ab-initio* phase analysis. Unrelated to applications in electron diffraction, the reconstruction technique used in this study provides new ways to store image data and might offer new insight into 3D reconstruction problems related to medical imaging, preservation of ancient architecture and antiques, etc.

Overall, the current methodology to reconstruct spherical Kikuchi maps from experimental patterns onto a simulated Kikuchi sphere still contains a few limitations. First, some of the edges of different experimental patterns are not perfectly aligned on the Kikuchi sphere, which could be attributed to uncertainties in the experimental parameters such as microscope/detector alignment (systematic misalignment), Hough indexing (random misalignment), pattern center position (random/systematic misalignment) and detector/sample tilt angle (systematic misalignment) or lattice strain (random misalignment) that changes the interplanar anglesDecoupling all the random or systematic misalignment would indeed improve the alignment of the images, although it is in practice a very challenging task. In this study, the normalized cross-correlation method between kinematical simulation and experimental pattern is used to refine pattern selection method, which helps to look for patterns that are more aligned with its corresponding kinematical simulation. Nevertheless, future improvement in this approach should incorporate dynamical simulation to enhance the effectiveness of the cross-correlation. This limitation in our study also motivates alternative geometry for doing EBSD e.g. use of tilt-free EBSD to reduce the uncertainties in setting up the scan on carefully heat-treated polycrystalline samples.  Another alternative solution to improve the misalignment issue would be to use a single crystal sample that is physically rotated with respect to the detector screen to collect diffraction patterns of different orientations. Nevertheless, small misalignment due to Hough-based orientation measurement will still be present because of angular errors.  Second, the latter EBSP added to the reconstructed map simply overwrites the intensity values of the previously mapped one in the overlapping region without any post-processing. An important feature about Kikuchi band profiles, as discussed in Day's paper, is the asymmetry of intensity distribution i.e. excess and deficiency lines, which is related to the anisotropy scattering of the inelastically scattered electrons in EBSD (Winkelmann, 2008). This asymmetry effect has not been accounted for in this study. Future development of an appropriate method to render averaged intensity values from overlapping regions along with edge stitching algorithm might improve the quality of spherical Kikuchi maps, correct the asymmetric Kikuchi band profiles and even assist in solving the alignment issue mentioned above.  Third, the method does not allow for the reconstruction of an unknown phase at this point, which could be made possible in the future using an image stitching algorithm based on overlapping features between different collected patterns.

4 Conclusion

We present here a novel approach to reconstruct spherical Kikuchi maps from experimentally collected and simulated electron backscatter diffraction patterns by overlaying them, in an automated way, onto a simulated Kikuchi sphere. This work demonstrates the feasibility of reconstructing a spherical Kikuchi map of a given phase based on any set of experimental patterns, an idea suggested in 2008. This method consists of the following steps: 1) pattern selection based on multiple threshold values; 2) extraction of multiple experimental parameters; 3) the generation of a kinematically simulated Kikuchi sphere as a 'skeleton'; and 4) overlaying the inverse gnomonic projection of multiple selected patterns without interpolating the intensity values, after appropriate pattern center calibration and refinement. We demonstrate the ability to reconstruct more than 90% of the Kikuchi sphere using as few as 7 patterns. Lastly, we discuss some of the applications and opportunities for potential future refinement of the technique. The demonstrated method for reconstructing 3D Kikuchi maps from EBSD patterns is likely to be an important tool for developing and improving new indexing algorithms, pattern center refinement, or phase identification.

5 Acknowledgment

C. Zhu would like to acknowledge helpful information provided by Michael Hjelmstad from Oxford Instruments. C. Zhu would like to acknowledge the Dissertation Year Fellowship provided by the home department. K. Kaufmann was supported by the Department of Defense (DoD) through the National Defense Science and Engineering Graduate Fellowship (NDSEG) Program. K. Kaufmann would like to acknowledge the generous support of the ARCS Foundation, San Diego Chapter.

of Multivariate Statistical Methods and Simulation Libraries to Analysis of Electron Backscatter Diffraction and Transmission Kikuchi Diffraction Datasets. *Ultramicroscopy* **196**, 88–98.

WILKINSON, A. J., MEADEN, G. & DINGLEY, D. J. (2006). High-resolution elastic strain measurement from electron backscatter diffraction patterns: New levels of sensitivity. *Ultramicroscopy* **106**, 307–313.

WILKINSON, A. J., MOLDOVAN, G., BRITTON, T. B., BEWICK, A., CLOUGH, R. & KIRKLAND, A. I. (2013). Direct detection of electron backscatter diffraction patterns. *Phys. Rev. Lett.* **111**, 065506.

WILLIAMS, D. B. & CARTER, C. B. (2009). *Transmission electron microscopy: A textbook for materials science*. Springer.

WINKELMANN, A. (2008). Dynamical effects of anisotropic inelastic scattering in electron backscatter diffraction. *Ultramicroscopy* **108**, 1546–1550.

WINKELMANN, A. (2010). Principles of depth-resolved Kikuchi pattern simulation for electron backscatter diffraction. *J. Microsc.* **239**, 32–45.

WINKELMANN, A., NOLZE, G., VOS, M., SALVAT-PUJOL, F. & WERNER, W. S. M. (2016). Physics-based simulation models for EBSD: Advances and challenges. *IOP Conf. Ser. Mater. Sci. Eng.* **239**, 32–45.

WRIGHT, S. I. & ADAMS, B. L. (1992). Automatic analysis of electron backscatter diffraction patterns. *Metall. Trans. A* **23**, 759–767.37

in high-purity titanium. *Int. J. Plast.*

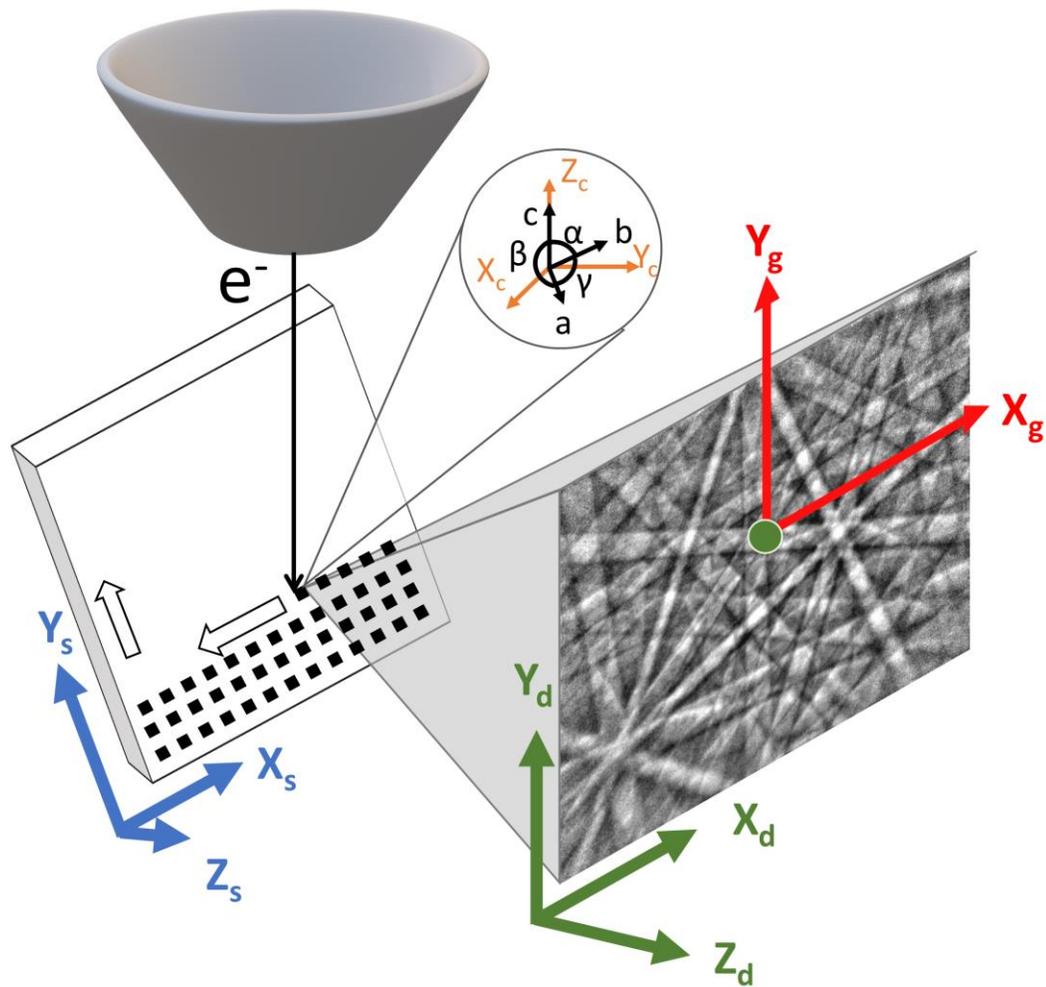

Fig. 1. Schematic of EBSD pattern collection showing all relevant coordinate systems: crystal lattice frame (black), cartesian crystal frame (orange), sample frame (blue), detector frame (green) and gnomonic projection frame (red).

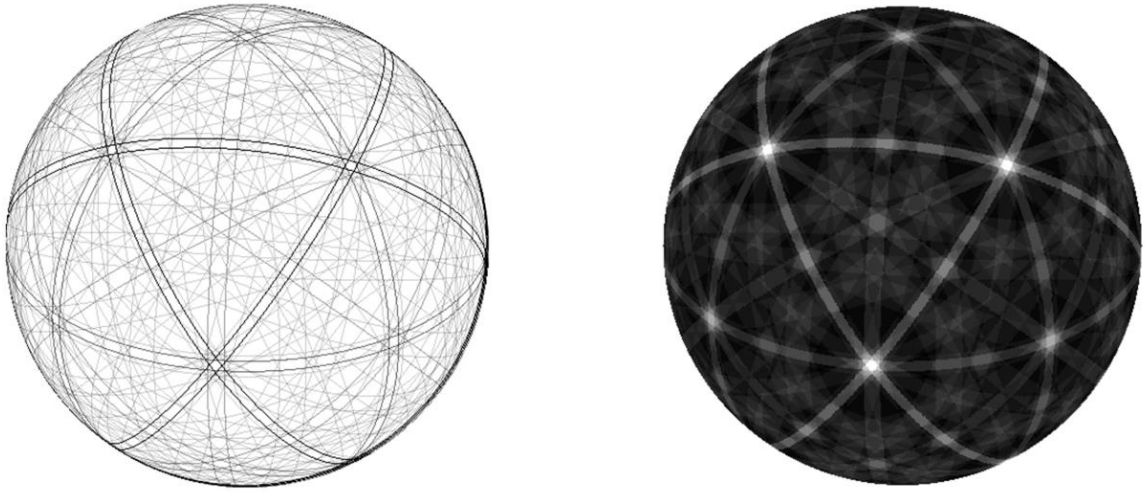

Fig. 2. Kinematically simulated Kikuchi sphere for aluminum at an accelerating voltage of 20kV: (left) spherical Kikuchi lines with its intensity scaled by $|F_{hkl}|$; (right) kinematical Kikuchi sphere with its band intensity scaled by $|F_{hkl}|$.

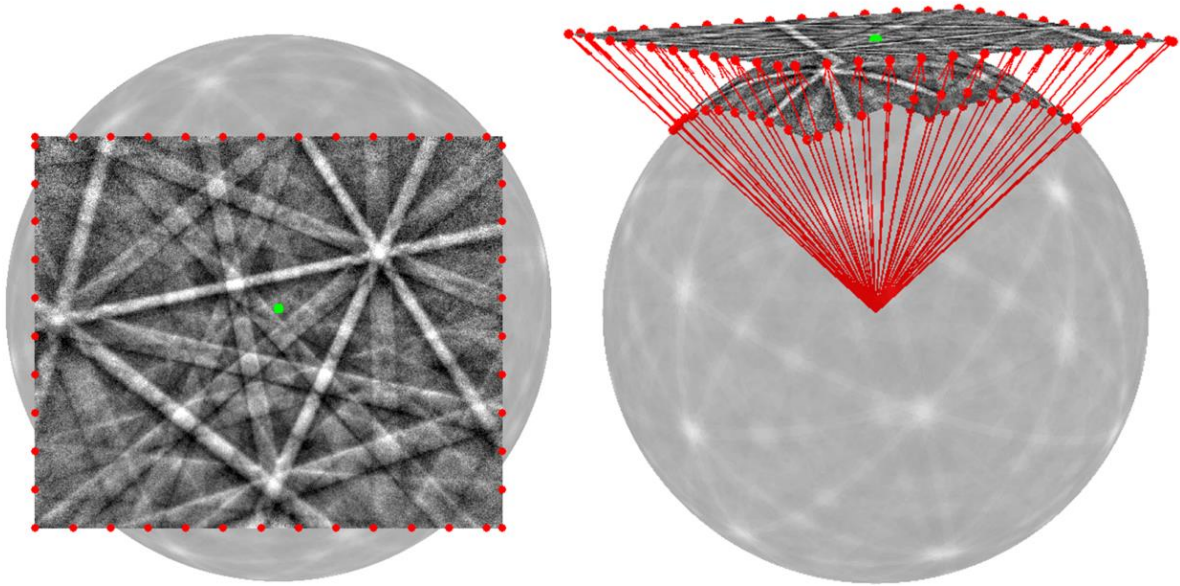

Fig. 3. Schematic showing the gnomonic projection (red arrows) of an as-received experimental Kikuchi pattern, collected from an aluminum (fcc) sample, viewed (left) on the axis of the detector and (right) viewed from the side. The experimental pattern is tangent to the simulated Kikuchi sphere, which intersects at the green point.

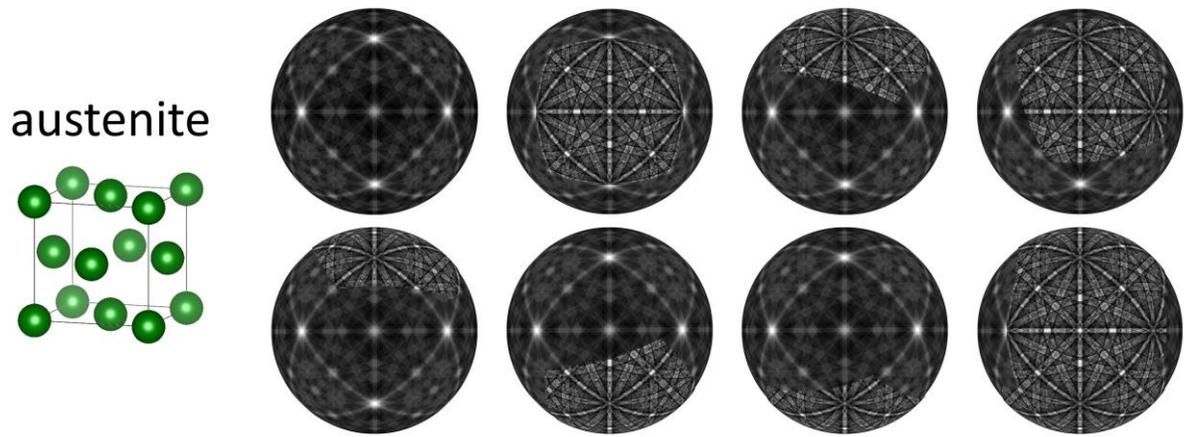

Fig. 4. (left) Crystal unit cells of austenite (fcc, a=b=c=3.65 Å, $\alpha=\beta=\gamma=90°$) (right) Automatic partial reconstruction process of a single spherical Kikuchi map for austenite from dynamically simulated patterns overlaid on a kinematically simulated austenite Kikuchi sphere at 20kV.

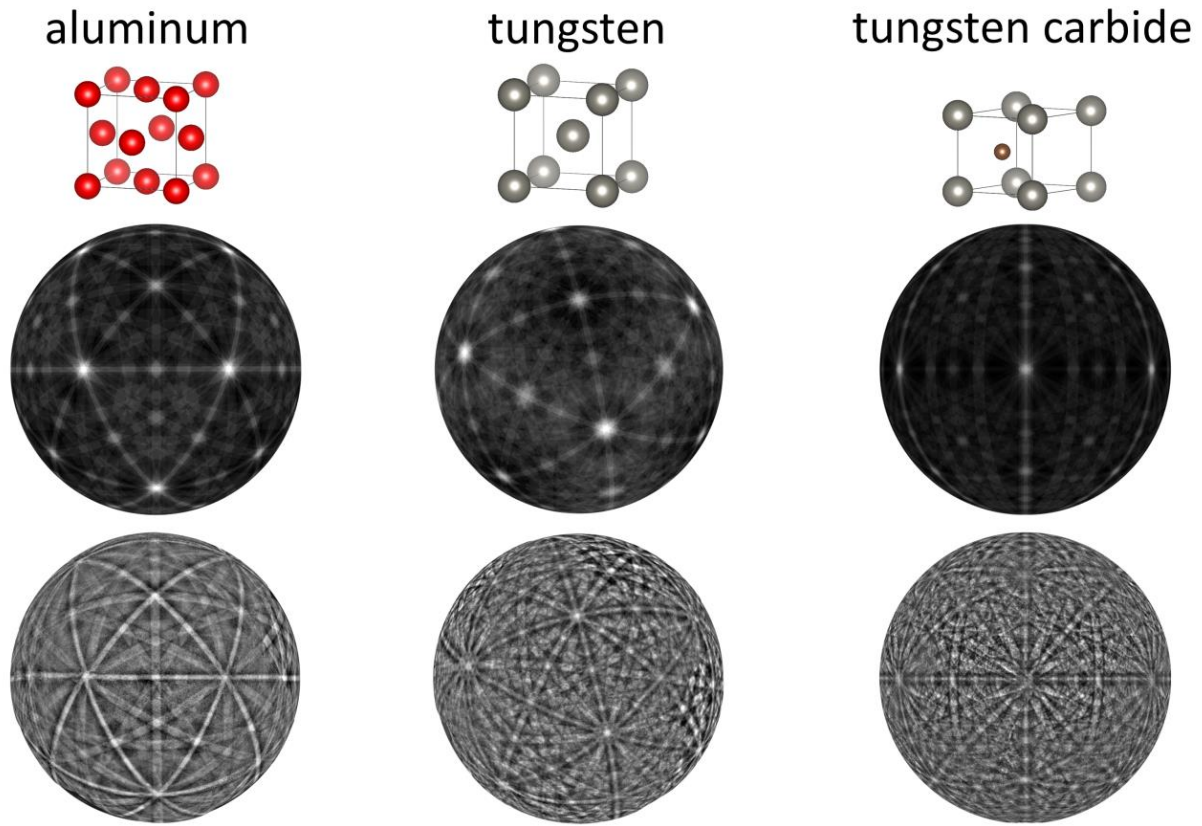

Fig. 5. (Top row from left to right) Crystal unit cells of aluminum (fcc, a=b=c=4.04 Å, $\alpha=\beta=\gamma=90°$), tungsten (bcc, a=b=c=3.19 Å, $\alpha=\beta=\gamma=90°$) and tungsten carbide (hexagonal, a=b=2.93 Å, c=2.85 Å, $\alpha=\beta=90°$, $\gamma=120°$); (middle row from left to right) kinematically simulated Kikuchi sphere for aluminum, tungsten and tungsten carbide at 20 kV; (bottom row from left to right) A partially reconstructed spherical Kikuchi map from experimental patterns overlaid on a simulated Kikuchi sphere for aluminum, tungsten, and tungsten carbide at 20kV. The kinematically simulated sphere is utilized as visual confirmation to the reader that the experimentally collected EBSPs are being mapped to the correct location on the sphere.

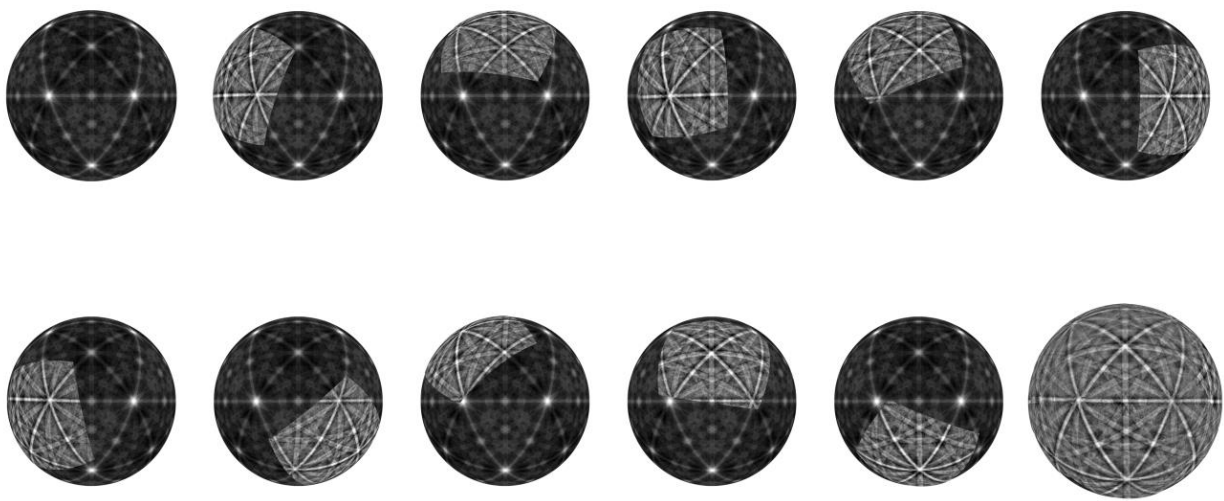

S.1: The automatic reconstruction process of a spherical Kikuchi map for aluminum from high-resolution experimental patterns overlaid on a kinematically simulated aluminum Kikuchi sphere at 20kV. The simulated sphere underlay is left in view as visual confirmation to the reader that the experimentally collected EBSPs are being mapped to the correct location.

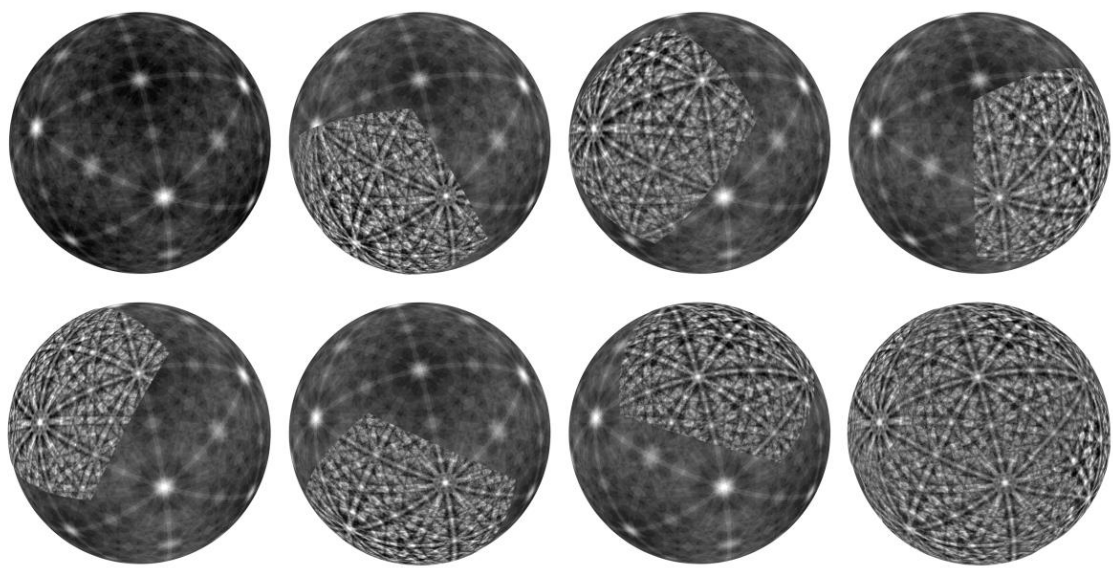

S.2: The automatic reconstruction process of a spherical Kikuchi map for tungsten from high-resolution experimental patterns overlaid on a kinematically simulated tungsten Kikuchi sphere at 20kV. The simulated sphere underlay is left in view as visual confirmation to the reader that the experimentally collected EBSPs are being mapped to the correct location.

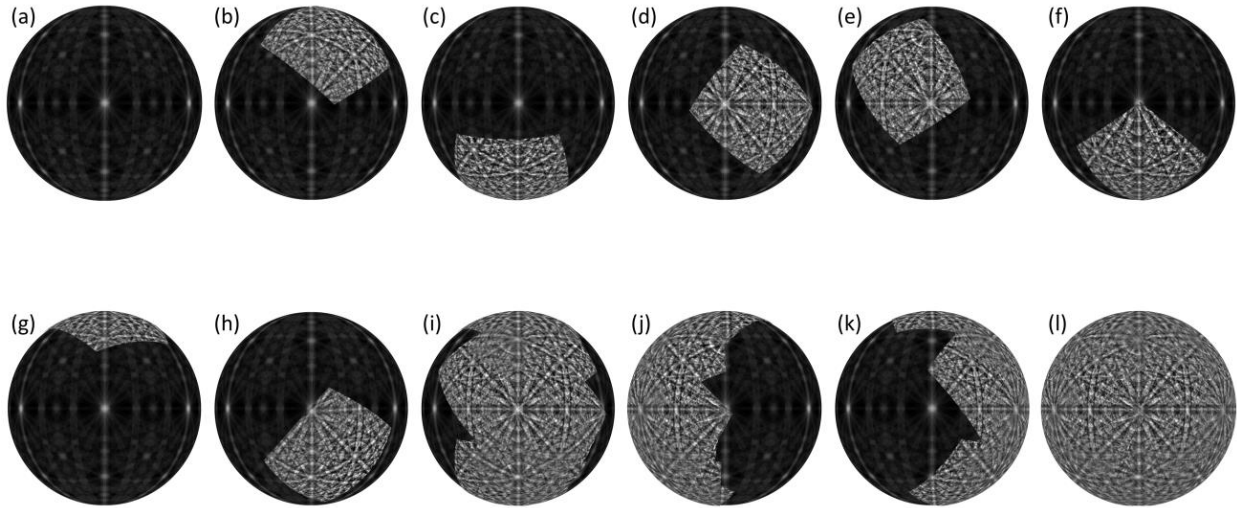

S.3: The automatic reconstruction process (a-l) of a spherical Kikuchi map for tungsten carbide from high-resolution experimental patterns overlaid on a kinematically simulated tungsten carbide Kikuchi sphere at 20kV. The simulated sphere underlay is left in view as visual confirmation to the reader that the experimentally collected EBSPs are being mapped to the correct location. (i) represents original set ($[\phi_1, \Phi, \phi_2]$) of reconstructed patterns. (j) represents the rotation ($[\phi_1, \Phi, \phi_2 + 60]$) of the original set by +60°. (k) represents the rotation ($[\phi_1, \Phi, \phi_2 - 60]$) of the original set by -60°. (l) combines (i-k) into fully reconstructed hemisphere of the spherical Kikuchi map.